\documentstyle[amssymb,preprint,aps]{revtex}

\begin{document}
\title{INTERFERENCE PHENOMENA IN SPONTANEOUS EMISSION FROM DRIVEN MULTILEVEL ATOMS}
\author{V.O.Chaltykyan, A.D. Gazazyan, and Y.T.Pashayan}
\address{Institute for Physical Research, \\
Armenian National Academy of Sciences, \\
Ashtarak-2, 378410, Armenia\\
(email: alfred@ipr.sci.am)}
\date{}
\maketitle

\begin{abstract}
We study spontaneous emission from an atom under the action of laser fields.
We consider two different energy level diagrams. The first one consists of
two levels resonantly driven by laser radiation where either of levels may
decay to a separate level. For such a system we show that the presence of
the second decay channel may deteriorate the destructive interference
occurring in case of one decay channel because of Autler-Townes effect. The
second diagram represents two two-level resonantly driven systems with the
upper levels decaying to a common level. For this diagram we show that there
is no interference between the two decay channels when the laser fields are
described in the Fock representation, while in case of definite-phase
classical fields such interference takes place and is partially or
completely destructive or constructive depending on the initial conditions
and on the mutual orientation of the spontaneous emission dipole moments.
\end{abstract}

\section{Introduction}

The investigations of the decay of a quantum-mechanical system affected by
an external electromagnetic field and the interference phenomena between the
different decay channels are of fundamental as well as practical interest.
All the quantum processes under the action of electromagnetic radiation can
be divided into two groups: i) processes occurring also in the absence of an
externally applied field and affected by this field and ii) processes that
are stimulated by the electromagnetic field and do not take place when there
is no radiation. In this sense, spontaneous transitions from excited states
of atoms pertain to the first group, while the excitation of an atom by an
external electromagnetic field with subsequent spontaneous transition to a
lower state pertains to the second group.

The spontaneous emission spectrum from an isolated atomic level is known to
be described by the Lorentz profile.The effect of laser radiation leads to a
significant change in both the lifetime and spectral line shape of
spontaneous emission. In particular, it is known that the decay law of an
atomic level depends upon the state of the external quantized
electromagnetic field \cite{Alimov}. In a multilevel system, transitions
between the states make the spectrum of spontaneous emission more
complicated and dependent upon the radiation intensity.

Recently much attention has been focused on the study of spontaneous
emission spectrum of an atom. Of particular interest is the quantum Zeno
effect, i.e., retardation (or complete stopping) of the decay of a quantum
system during periodic ''observations'' of this system. Such an
''observation'' is realized, in particular, by periodic excitation with
laser pulses. However, an essential modification of spontaneous emission
spectrum can also be achieved by a cw laser, in particular, interference
between the different decay channels in an atom can result in appearance of
dark lines in the spontaneous emission spectrum if interference is
destructive.The work \cite{Zhu}considered an atom with two excited levels
coupled by external electromagnetic filed and spontaneous transitions from
one of these levels. Quantum interference in the spontaneous emission of a
four-level atom is considered in \cite{Imamoglu} for the case where two
near-lying levels undergoing a spontaneous transition to a lower level are
excited by a laser radiation from some other level. The possibility of
reducing spontaneous emission due to quantum interference is studied.
Spontaneous emission and the modification of the lifetime of a level is
investigated in \cite{Facchi}.

Resonance fluorescence of an atom in a high-Q resonator when the atomic
transition is coupled to the ''teilored'' vacuum, as well as spontaneous
emission from atoms in which the two transitions are coupled to different,
Markovian and non-Markovian, reservoirs, are studied in \cite{Paspalakis}.
It is shown that the appropriate choice of the mode density of the
non-Markovian reservoir essentially affects spontaneous emission in the
Markovian reservoir (free space vacuum). Ref.\cite{PK} considers the probe
absorption spectrum of a $\Lambda $-system when one of the atomic levels can
decay spontaneously near the edge of the photonic band gap and shows that
the probe is not absorbed at this frequency under certain conditions. In
Refs.\cite{PKK} it is proposed to use the phase difference of two coherent
fields for controlling spontaneous emission spectra.

In the present work we consider the influence of resonant laser radiation on
the spontaneous emission spectra in four- and five-level atoms (Figs. 1a,b).
We first obtain the wave functions of the driven atoms, as well as the wave
function of spontaneous photons in frames of the Wigner-Weisskopf theory 
\cite{Weisskopf}. In a four-level atomic system (Fig. 1a) the two upper
levels are coupled by laser field and emit spontaneous photons when decaying
to different states. The work\cite{Zhu} considered just the system of the
Fig.1a without the fourth level and showed that destructive or constructive
interference may occur, depending on initial conditions, between the
transitions from the components of the Autler-Townes doublet formed in laser
field. In case of destructive interference a dark line appears in the
fluorescence spectrum. We investigate the influence of the additional decay
channel (to the fourth level) on this spectrum and show, in particular, that
the dark line disappears generally, but may be observed under certain
conditions.

We then consider a five-level atomic system (Fig. 1b) where the two upper
levels 2 and 4 are excited by two laser fields from the lower states 1 and 3
with subsequent spontaneous transition to the final state $f$. We will
obtain that if the driving laser fields are described in the photon number
representation, the system behaves like two independent two-level systems,
while in definite-phase classical laser fields interference takes place
between the different decay channels. We will examine the character of this
interference and its results for different initial conditions and values of
relevant parameters.

\section{Four-level atomic system}

Consider a four-level atom in the field of quanyized laser radiation at the
frequency $\omega $ close to resonance with the transition $1\rightarrow 2$
(Fig.1a). Under the action of the laser field a coupled ''atom+field'' state
of levels $1$ and $2$ is formed. Let us assume that the atom is initially in
a superposition of these states and follow the evolution of the system, when
spontaneous transitions $1\rightarrow 3$ and $2\rightarrow 4$ are possible.

The full Hamiltonian of the system reads

\begin{equation}
\begin{array}{c}
H=H_{at}+\omega c^{+}c+\beta ^{+}c+c^{+}\beta +\sum\limits_{{\bf k},\lambda
}\omega _{{\bf k}}c_{{\bf k},\lambda }^{+}c_{{\bf k},\lambda }+ \\ 
+\sum\limits_{{\bf k},\lambda }\left[ \beta ^{+}({\bf k},\lambda )c_{{\bf k}%
,\lambda }+c_{{\bf k},\lambda }^{+}\beta ({\bf k},\lambda )\right] ,
\end{array}
\label{2_1}
\end{equation}
where $H_{at}$\ is the Hamiltonian of the bare atom

\begin{equation}
H_{at}\left| i\right\rangle =\varepsilon _{i}\left| i\right\rangle ,\qquad
(i=1,2,3,4),  \label{2_2}
\end{equation}
$\beta ^{+}$ and $\beta $ are the operators of dipole transitions from state
1 to state 2 and back ($\beta =\sqrt{2\pi \omega /\hbar V}\left( {\bf ed}%
\right) $, ${\bf e}$ being the polarization unit vector of laser field and
other notation commonly used) under the action of the external field of
frequency $\omega $, $\beta ^{+}({\bf k},\lambda )$ and $\beta ({\bf k}%
,\lambda )$ are similar operators for spontaneous photons with the momentum $%
{\bf k}$ and polarization $\lambda $; $c^{+},c,c_{{\bf k},\lambda }^{+}$ and 
$c_{{\bf k},\lambda }$ are the operators of annihilation and creation of
laser field and spontaneous photons, respectively.

The solution to the Shr\"{o}edinger equation with the Hamiltonian (\ref{2_1}%
) can be in the interaction representation written as

\begin{equation}
\begin{array}{c}
\Phi (t)=C_{1}(t)\left| 1,n,0\right\rangle +C_{2}(t)\left|
2,n-1,0\right\rangle + \\ 
+\sum\limits_{{\bf k},\lambda }C_{3{\bf k},\lambda }\left| 3,n,1_{{\bf k}%
,\lambda }\right\rangle +\sum\limits_{{\bf k},\lambda }C_{4{\bf k},\lambda
}\left| 4,n-1,1_{{\bf k},\lambda }\right\rangle
\end{array}
,  \label{2_4}
\end{equation}
where $\left| i,n,l\right\rangle $ stands for the state of the system with
the atom in $i$-th level, $n$ photons in the laser field, and $l$
spontaneously emitted photons.

By\ substituting this expression into the Shr\"{o}edinger equation we obtain
a set of coupled differential equations for the expansion coefficients,

\begin{equation}
\begin{array}{c}
i%
{\displaystyle{dC_{1}(t) \over dt}}%
=\sqrt{n}\beta e^{-i\nu t}C_{2}(t)+\sum\limits_{{\bf k},\lambda }\beta
_{13}^{+}({\bf k},\lambda )e^{i(\varepsilon _{1}-\varepsilon _{3}-\omega _{%
{\bf k}})t}C_{3{\bf k},\lambda }(t) \\ 
i%
{\displaystyle{dC_{2}(t) \over dt}}%
=\sqrt{n}\beta ^{+}e^{i\nu t}C_{1}(t)+\sum\limits_{{\bf k},\lambda }\beta
_{24}^{+}({\bf k},\lambda )e^{i(\varepsilon _{2}-\varepsilon _{4}-\omega _{%
{\bf k}})t}C_{4{\bf k},\lambda }(t) \\ 
i%
{\displaystyle{dC_{3{\bf k},\lambda }(t) \over dt}}%
=\beta _{31}({\bf k},\lambda )e^{-i(\varepsilon _{1}-\varepsilon _{3}-\omega
_{{\bf k}})t}C_{1}(t) \\ 
i%
{\displaystyle{dC_{4{\bf k},\lambda }(t) \over dt}}%
=\beta _{42}({\bf k},\lambda )e^{-i(\varepsilon _{2}-\varepsilon _{4}-\omega
_{{\bf k}})t}C_{2}(t),
\end{array}
\label{dif_eqs}
\end{equation}
where $\nu $ is the detuning of resonance ($\nu =\varepsilon
_{2}-\varepsilon _{1}-\omega $).

By eliminating $C_{3,4{\bf k},\lambda }(t)$ (with the initial conditions $%
C_{3,4{\bf k},\lambda }(0)=0$ ) we obtain

\begin{equation}
\begin{array}{c}
{\displaystyle{dC_{1}(t) \over dt}}%
=-i\sqrt{n}\beta e^{-i\nu t}C_{2}(t)-\sum\limits_{{\bf k},\lambda }\left|
\beta _{13}({\bf k},\lambda )\right| ^{2}\int\limits_{0}^{t}e^{i(\varepsilon
_{1}-\varepsilon _{3}-\omega _{{\bf k}})(t-t^{\prime })}C_{1}(t^{\prime
})dt^{\prime } \\ 
{\displaystyle{dC_{2}(t) \over dt}}%
=-i\sqrt{n}\beta ^{\ast }e^{i\nu t}C_{1}(t)-\sum\limits_{\overrightarrow{k}%
,\lambda }\left| \beta _{24}({\bf k},\lambda )\right|
^{2}\int\limits_{0}^{t}e^{i(\varepsilon _{2}-\varepsilon _{4}-\omega _{{\bf k%
}})(t-t^{\prime })}C_{2}(t^{\prime })dt^{\prime }
\end{array}
.
\end{equation}

In further calculations we will use the Wigner-Weiskopf approximation\cite
{Weisskopf}. In this case, by means of the Laplace transformation technique
we obtain the following solutions to the Eqs.(5):

\[
\begin{array}{c}
C_{1}(t)=\frac{1}{\Omega }e^{-\frac{1}{4}(\Gamma _{1}+\Gamma _{2})t}e^{-%
\frac{i}{2}(\nu +\Delta _{1}+\Delta _{2})t}\times \\ 
\times \left\{ \left[ \Omega \cos \frac{\Omega }{2}t+i\left( \nu -\Delta
_{1}+\Delta _{2}+\frac{i}{2}(\Gamma _{1}-\Gamma _{2})\right) \sin \frac{%
\Omega }{2}t\right] C_{1}-2i\sqrt{n}\beta C_{2}\sin \frac{\Omega }{2}%
t\right\}
\end{array}
\]

\begin{equation}
\begin{array}{c}
C_{2}(t)=\frac{1}{\Omega }e^{-\frac{1}{4}(\Gamma _{1}+\Gamma _{2})t}e^{\frac{%
i}{2}(\nu -\Delta _{1}-\Delta _{2})t}\times \\ 
\times \left\{ \left[ \Omega \cos \frac{\Omega }{2}t-i\left( \nu -\Delta
_{1}+\Delta _{2}+\frac{i}{2}(\Gamma _{1}-\Gamma _{2})\right) \sin \frac{%
\Omega }{2}t\right] C_{2}+2i\sqrt{n}\beta ^{\ast }C_{1}\sin \frac{\Omega }{2}%
t\right\}
\end{array}
,  \label{2_12}
\end{equation}

\bigskip where $C_{1}=C_{1}(0)$, $C_{2}=C_{2}(0)$ and the shifts and widths
of corresponding levels, and the complex Rabi frequency are defined as (the
notation $P$ stands here and below for the principal value)

\begin{equation}
\begin{array}{c}
\Delta _{1}=-P\sum\limits_{{\bf k},\lambda }%
{\displaystyle{\left| \beta _{13}({\bf k},\lambda )\right| ^{2} \over \omega _{{\bf k}}-\varepsilon _{1}+\varepsilon _{3}}}%
,\qquad \Gamma _{1}=2\pi \sum\limits_{{\bf k},\lambda }\left| \beta _{13}(%
{\bf k},\lambda )\right| ^{2}\delta (\omega _{{\bf k}}-\varepsilon
_{1}+\varepsilon _{3}) \\ 
\Delta _{2}=-P\sum\limits_{{\bf k},\lambda }%
{\displaystyle{\left| \beta _{24}({\bf k},\lambda )\right| ^{2} \over \omega _{{\bf k}}-\varepsilon _{2}+\varepsilon _{4}}}%
,\qquad \Gamma _{2}=2\pi \sum\limits_{{\bf k},\lambda }\left| \beta _{24}(%
{\bf k},\lambda )\right| ^{2}\delta (\omega _{{\bf k}}-\varepsilon
_{2}+\varepsilon _{4}).
\end{array}
\label{delta_gamma}
\end{equation}

$\ \ \ \ \ \ \ \ \ \ \ \ \ \ \ \ \ \ \ \ \ \ \ \ \ \ \ \ \ \ \ \ \ \ \ \ \ \
\ \ \Omega =\sqrt{\left[ \nu -\Delta _{1}+\Delta _{2}+\frac{i}{2}(\Gamma
_{1}-\Gamma _{2})\right] ^{2}+4n\left| \beta \right| ^{2}}.$

By using the expressions (6) in the second pair of Eqs.(4), we finally obtain

\[
\begin{array}{c}
C_{3{\bf k},\lambda }(t)=\frac{\beta _{31}({\bf k},\lambda )}{2\Omega }\times
\\ 
\times 
\begin{array}{c}
\begin{array}{c}
\{\left[ \left( \nu -\Delta _{1}+\Delta _{2}+\frac{i}{2}(\Gamma _{1}-\Gamma
_{2})+\Omega \right) C_{1}-2\sqrt{n}\beta C_{2}\right] \times \\ 
\times 
{\displaystyle{e^{-i\left[ \nu _{1{\bf k}}+\frac{1}{2}\left( \nu +\Delta _{1}+\Delta _{2}-\frac{i}{2}(\Gamma _{1}+\Gamma _{2})-\Omega \right) \right] t}-1 \over \nu _{1{\bf k}}+\frac{1}{2}\left( \nu +\Delta _{1}+\Delta _{2}-\frac{i}{2}(\Gamma _{1}+\Gamma _{2})-\Omega \right) }}%
-
\end{array}
\\ 
-\left[ \left( \nu -\Delta _{1}+\Delta _{2}+\frac{i}{2}(\Gamma _{1}-\Gamma
_{2})-\Omega \right) C_{1}-2\sqrt{n}\beta C_{2}\right] \times \\ 
\times 
{\displaystyle{e^{-i\left[ \nu _{1{\bf k}}+\frac{1}{2}\left( \nu +\Delta _{1}+\Delta _{2}-\frac{i}{2}(\Gamma _{1}+\Gamma _{2})+\Omega \right) \right] t}-1 \over \nu _{1{\bf k}}+\frac{1}{2}\left( \nu +\Delta _{1}+\Delta _{2}-\frac{i}{2}(\Gamma _{1}+\Gamma _{2})+\Omega \right) }}%
\}
\end{array}
\end{array}
\]

\begin{equation}
\begin{array}{c}
C_{{\bf k},\lambda }(t)=-\frac{\beta _{42}({\bf k},\lambda )}{2\Omega }\times
\\ 
\times 
\begin{array}{c}
\{\left[ \left( \nu -\Delta _{1}+\Delta _{2}+\frac{i}{2}(\Gamma _{1}-\Gamma
_{2})-\Omega \right) C_{2}-2\sqrt{n}\beta ^{\ast }C_{1}\right] \times \\ 
\times 
{\displaystyle{e^{-i\left[ \nu _{2{\bf k}}-\frac{1}{2}\left( \nu -\Delta _{1}-\Delta _{2}+\frac{i}{2}(\Gamma _{1}+\Gamma _{2})+\Omega \right) \right] t}-1 \over \nu _{2{\bf k}}-\frac{1}{2}\left( \nu -\Delta _{1}-\Delta _{2}+\frac{i}{2}(\Gamma _{1}+\Gamma _{2})+\Omega \right) }}%
- \\ 
-\left[ \left( \nu -\Delta _{1}+\Delta _{2}+\frac{i}{2}(\Gamma _{1}-\Gamma
_{2})+\Omega \right) C_{2}-2\sqrt{n}\beta ^{\ast }C_{1}\right] \times \\ 
\times 
{\displaystyle{e^{-i\left[ \nu _{2{\bf k}}-\frac{1}{2}\left( \nu -\Delta _{1}-\Delta _{2}+\frac{i}{2}(\Gamma _{1}+\Gamma _{2})-\Omega \right) \right] t}-1 \over \nu _{2{\bf k}}-\frac{1}{2}\left( \nu -\Delta _{1}-\Delta _{2}+\frac{i}{2}(\Gamma _{1}+\Gamma _{2})-\Omega \right) }}%
\}
\end{array}
,
\end{array}
\label{2_14}
\end{equation}
with the detunings $\nu _{1{\bf k}}=\varepsilon _{1}-\varepsilon _{3}-\omega
_{{\bf k}},$ $\nu _{2{\bf k}}=\varepsilon _{2}-\varepsilon _{4}-\omega _{%
{\bf k}}.$ Expressions (6) and (8) determine the wave function (3) which
allows for spontaneous transitions.

It is seen from the expressions for $C_{1}(t)$ and $C_{2}(t)$ that these
coefficients tend to zero as $t\rightarrow \infty $: $C_{1}(\infty
)=C_{2}(\infty )=0$ . The wave function of the system $\Phi (t)$ will then
have the form

\begin{equation}
\Phi (\infty )=\left| ph_{1}\right\rangle \left| 3,n\right\rangle +\left|
ph_{2}\right\rangle \left| 4,n-1\right\rangle ,  \label{2_16}
\end{equation}
where

\begin{equation}
\begin{array}{c}
\left| ph_{1}\right\rangle =\sum\limits_{{\bf k},\lambda }C_{3{\bf k}%
,\lambda }(\infty )\left| 1_{{\bf k}\lambda }\right\rangle \\ 
\left| ph_{2}\right\rangle =\sum\limits_{{\bf k},\lambda }C_{4{\bf k}%
,\lambda }(\infty )\left| 1_{{\bf k}\lambda }\right\rangle
\end{array}
\label{2_17}
\end{equation}
are the wave functions of spontaneous photons emitted in the transitions $%
1\rightarrow 3$ and $2\rightarrow 4$, respectively.

From Eq. (\ref{2_16}) it follows that there is no interference between the
decay channels to levels $3$ and $4$ because the final atomic states are
different. Now, the spontaneous emission spectrum may be obtained in terms
of $\left| C_{3{\bf k},\lambda }(\infty )\right| ^{2}+\left| C_{4{\bf k}%
,\lambda }(\infty )\right| ^{2}$, which after summation over polarizations
of emitted photons and integration over the angles gives $I$\ $(\omega
_{k})=\omega _{k}^{3}S(\omega _{k})$, where

\begin{equation}
\begin{array}{c}
{\Large S(\omega }_{k}{\Large )\varpropto }\left| {\bf d}_{31}\right| ^{2}%
{\displaystyle{(\varepsilon _{2}-\varepsilon _{3}-\omega -\omega _{k}+\Delta _{2}-\frac{i\Gamma _{2}}{2})C_{1}-C_{2}\sqrt{n}\beta  \overwithdelims|| (\varepsilon _{1}-\varepsilon _{3}-\omega _{k}+\Delta _{1}-\frac{i\Gamma _{1}}{2})(\varepsilon _{2}-\varepsilon _{3}-\omega -\omega _{k}+\Delta _{2}-\frac{i\Gamma _{2}}{2})-n\left| \beta \right| ^{2}}}%
^{2}{\Large +} \\ 
{\Large +}\left| {\bf d}_{42}\right| ^{2}%
{\displaystyle{(\varepsilon _{1}-\varepsilon _{4}+\omega -\omega _{k}+\Delta _{1}-\frac{i\Gamma _{1}}{2})C_{2}+C_{1}\sqrt{n}\beta _{12}^{\ast } \overwithdelims|| (\varepsilon _{2}-\varepsilon _{4}-\omega _{k}+\Delta _{2}-\frac{i\Gamma _{2}}{2})(\varepsilon _{1}-\varepsilon _{4}+\omega -\omega _{k}+\Delta _{1}-\frac{i\Gamma _{1}}{2})-n\left| \beta \right| ^{2}}}%
^{2}.
\end{array}
\label{2_22}
\end{equation}

This is the basic result of this section; the expression (11) determines the
fluorescence spectrum of the four-level atom under study.

We will analize this spectrum in two special cases where i) atom is
initially in the state 1 ($C_{1}=1,C_{2}=0$) and ii) the atom is initially
prepared in the symmetric superposition state ($C_{1}=C_{2}=1/\sqrt{2}$).

The system of Fig.1a in the absence of the level 4 has been considered in
[2] where it was shown that if the atom is initially in the state 1(2), in
the fluorescence spectrum a dark(bright) line appears; the phenomenon was
explained as a result of destructive(constructive) interference between the
splitted laser-driven levels. This result is contained as a special case in
the formula (11). Indeed, if we substitute $C_{1}=1,C_{2}=0$ into (11) and
assume the absence of level 4 ($d_{42}=0,$ $\Delta _{2}=0,$ $\Gamma _{2}=0$%
), a dark state appears in the spectrum at the frequency $\omega
_{k}=\varepsilon _{2}-\varepsilon _{3}-\omega $. But when the level 4 is
present providing another decay channel, the dark line is absent. This case
is shown in Fig.2. The figure demonstrates the modification of the
fluorescence spectrum at growing value of $\ \Omega $. At a small value
(0.5) a single peak is observed instead of a dark line. This peak
corresponds to mostly the transition 1 - 3, since the level 2 is coupled to
the level 1 weakly. At greater values of $\ \Omega $ other three peaks
appear corresponding to the splitting of the levels 1 and 2 due to laser
coupling (Autler-Townes effect).

Consider now the second case where the atom is initially in the
superposition of the states 1 and 2. By substituting $C_{1}=C_{2}=1/\sqrt{2}$
into (11) we can easily obtain that under the conditions

\begin{equation}
\begin{array}{c}
\varepsilon _{3}-\varepsilon _{4}+\omega =\nu \\ 
\Delta _{2}-\frac{i\Gamma _{2}}{2}=-\left( \Delta _{1}+\frac{i\Gamma _{1}}{2}%
\right) =\sqrt{n}\beta \qquad (\Gamma _{1}=\Gamma _{2},\quad \Delta
_{2}=-\Delta _{1})
\end{array}
\label{2_26}
\end{equation}
a dark line appears at the frequency $\omega _{k}=\varepsilon
_{2}-\varepsilon _{3}-\omega =\varepsilon _{1}-\varepsilon _{4}+\omega $ in
the fluorescence spectrum. This means that if the Lamb shifts of the levels
1 and 2 are equal and opposite directed, the widths of these levels are
equal, and the laser frequency is tuned to $((\varepsilon _{2}-\varepsilon
_{1})-(\varepsilon _{4}-\varepsilon _{3}))/2$,\ a complete destructive
interference takes place in each transition channel to levels $3$ and $4$
separately due to the Autler-Townes effect in the symmetric superposition
upper state.

\section{ Five-level atomic system}

Consider now the case of a five-level atomic system in the field of two
lasers at frequencies $\omega _{1}$ and $\omega _{2}$ (Fig.1b). The first
field of frequency $\omega _{1}$ is in resonance with the transition $%
1\rightarrow 2$, and the second field of frequency $\omega _{2}$ is in
resonance with the transition $3\rightarrow 4$. The atom initially ( $t=0$)
is supposed to be in a superposition of the states 1 and 3

\begin{equation}
\Phi (0)=C_{1}\left| 1\right\rangle +C_{3}\left| 3\right\rangle .
\label{3_1}
\end{equation}

Let us investigate the evolution of such a system when spontaneous
transitions from levels $2$ and $4$ to some intermediate level $f$ are
possible. If the driving laser fields will be described in the Fock
representation, the spontaneous transitions from states 2 and 4 to the level 
$f$ will result in the formation of a superposition of two orthogonal
states, $\left| f,n_{1}-1,n_{2},1_{{\bf k}\lambda }\right\rangle $ and $%
\left| f,n_{1},,n_{2}-1,1_{{\bf k}\lambda }\right\rangle $, where $n_{1}$
and $n_{2}$ are the photon numbers in the first and second laser fields,
respectively. In this case we will have two different independent processes
that do not interfere and the intensity of the spontaneous emission will be
the sum of intensities of separate spontaneous emissions in transitions $%
2\rightarrow f$ and $4\rightarrow f$.

However, the situation significantly changes when the upper levels 2and 4
are excited by classical fields with definite phases. In this case all the
levels participate in the process and the interference takes place between
different transition channels.

The Hamiltonian of the system in case of classical excitation has the form

\begin{equation}
\begin{array}{c}
H=H_{at}+V_{1}(t)+V_{2}(t)+ \\ 
+\sum\limits_{{\bf k},\lambda }\omega _{{\bf k}}c_{{\bf k}\lambda }^{+}c_{%
{\bf k}\lambda }+\sum\limits_{{\bf k},\lambda }\left( \beta ^{+}({\bf k}%
,\lambda )c_{{\bf k}\lambda }+c_{{\bf k}\lambda }^{+}\beta ({\bf k},\lambda
)\right)
\end{array}
,  \label{3_2}
\end{equation}
where $V_{1}(t)$ and $V_{2}(t)$ are the energies of the interaction of the
atom with the first and second lasers, respectively,

\begin{equation}
\begin{array}{c}
V_{1}(t)=V_{1}e^{-i\omega _{1}t}+V_{1}^{\ast }e^{i\omega _{1}t} \\ 
V_{2}(t)=V_{2}e^{-i\omega _{2}t}+V_{2}^{\ast }e^{i\omega _{2}t}
\end{array}
.
\end{equation}
The Hamiltonian of the bare atom has now the form

\begin{equation}
H_{at}=\sum_{j=1}^{4}\varepsilon _{j}\left| j\right\rangle \left\langle
j\right| +\varepsilon _{f}\left| f\right\rangle \left\langle f\right| .
\end{equation}

The solution to the Shroedinger equation with the Hamiltonian (14) can be
written as

\begin{equation}
\Phi (t)=\sum_{j=1}^{4}C_{j}(t)\left| j,0\right\rangle +\sum_{{\bf k}\lambda
}C_{f{\bf k}\lambda }(t)\left| f,1_{{\bf k}\lambda }\right\rangle .
\label{3_6}
\end{equation}

The expansion coefficients obey the following set of equations:

\begin{equation}
\begin{array}{c}
i%
{\displaystyle{dC_{1}(t) \over dt}}%
=V_{12}^{(1)}e^{-i\nu _{1}t}C_{2}(t) \\ 
i%
{\displaystyle{dC_{2}(t) \over dt}}%
=V_{12}^{(1)\ast }e^{i\nu _{1}t}C_{1}(t)+\sum\limits_{{\bf k},\lambda }\beta
_{f2}^{\ast }({\bf k},\lambda )e^{i\nu _{{\bf k}}^{(1)}t}C_{f{\bf k},\lambda
}(t) \\ 
i%
{\displaystyle{dC_{3}(t) \over dt}}%
=V_{34}^{(2)}e^{-i\nu _{2}t}C_{4}(t) \\ 
i%
{\displaystyle{dC_{4}(t) \over dt}}%
=V_{34}^{(2)\ast }e^{i\nu _{2}t}C_{3}(t)+\sum\limits_{{\bf k},\lambda }\beta
_{f4}^{\ast }({\bf k},\lambda )e^{i\nu _{{\bf k}}^{(2)}t}C_{f{\bf k},\lambda
}(t) \\ 
i%
{\displaystyle{dC_{f{\bf k,}\lambda }(t) \over dt}}%
=\beta _{f2}({\bf k},\lambda )e^{-i\nu _{{\bf k}}^{(1)}t}C_{2}(t)+\beta
_{f4}({\bf k},\lambda )e^{-i\nu _{{\bf k}}^{(2)}t}C_{4}(t),
\end{array}
\label{3_7}
\end{equation}
where

\begin{equation}
\begin{array}{c}
\nu _{1}=\varepsilon _{2}-\varepsilon _{1}-\omega _{1} \\ 
\nu _{2}=\varepsilon _{4}-\varepsilon _{3}-\omega _{2} \\ 
\nu _{{\bf k}}^{(1)}=\varepsilon _{2}-\varepsilon _{f}-\omega _{{\bf k}} \\ 
\nu _{{\bf k}}^{(2)}=\varepsilon _{4}-\varepsilon _{f}-\omega _{{\bf k}}
\end{array}
\end{equation}
are the detunings of the corresponding resonances.

Now, proceeding in the same way as in the preceeding section, we have

\begin{equation}
\ \Phi (\infty )=C_{1}(\infty )\left| 1,0\right\rangle +C_{3}(\infty )\left|
3,0\right\rangle +\left| ph\right\rangle \left| f\right\rangle .
\end{equation}

$\ \ \ \ \ \ \ \ \ \ \ \ \ \ \ \ \ \ \ \ \ \ \ \ \ \ \ \ \ \ \ \ \ \ \ \ \ $%
\ \ \ \ \ \ \ \ \ \ \ \ \ \ \ \ \ \ \ \ \ \ \ \ \ \ \ \ \ \ \ \ \ \ \ \ \ \
\ \ \ \ \ \ \ \ \ \ \ \ \ \ \ \ \ \ \ \ \ \ \ \ \ \ \ \ \ \ \ \ \ \ \ \ \ \
\ \ \ \ \ \ \ \ \ \ \ \ \ \ \ \ \ \ \ \ \ \ \ \ \ \ \ \ \ \ \ \ \ \ \ \ \ \
\ \ \ \ \ \ 

The last term in (20) again determines the spontaneous emission spectrum
which is proportional to $\left| C_{f{\bf k},\lambda }(\infty )\right| ^{2}$
and is obtained to be

$\ \ \ \ \ \ \ \ \ \ \ \ \ $\ \ \ \ \ \ \ \ \ \ \ \ \ \ \ \ \ \ \ \ \ \ \ \
\ \ \ \ \ \ \ \ \ \ \ \ \ \ \ \ \ \ \ \ \ \ \ \ \ \ \ \ \ \ \ \ \ \ \ \ \ \
\ \ \ \ \ \ \ \ \ \ \ \ \ \ \ \ \ \ \ \ 

\begin{equation}
\begin{array}{c}
I\ {\Large (\omega }_{k}{\Large )=\omega }_{k}^{3}{\Large S(\omega }_{k}%
{\Large )}; \\ 
{\Large S(\omega }_{k}){\Large \varpropto }{\bf d}_{f2}\sum\limits_{j=1}^{4}%
\frac{1}{\prod\limits_{l\neq j}^{4}(x_{j}-x_{l})}{\Large (}%
{\displaystyle{C_{1}V_{12}^{(1)\ast }\left[ (x_{j}-\varepsilon _{2}+\varepsilon _{4}-\nu _{2})(x_{j}-\varepsilon _{2}+\varepsilon _{4}+\Delta _{4}-\frac{i\Gamma _{4}}{2})-\left| V_{34}^{(2)}\right| ^{2}\right]  \over \omega _{k}-\varepsilon _{2}+\varepsilon _{f}-x_{j}}}%
+ \\ 
+%
{\displaystyle{C_{3}\frac{\sqrt{\Gamma _{2}\Gamma _{4}}}{2}(q+i)V_{34}^{(2)\ast }(x_{j}-\nu _{1}) \over \omega _{k}-\varepsilon _{2}+\varepsilon _{f}-x_{j}}}%
{\Large )}+ \\ 
\begin{array}{c}
{\Large +}{\bf d}_{f4}\sum\limits_{j=1}^{4}\frac{1}{\prod\limits_{l\neq
j}^{4}(\widetilde{x}_{j}-\widetilde{x}_{l})}{\Large (}%
{\displaystyle{C_{1}\frac{\sqrt{\Gamma _{2}\Gamma _{4}}}{2}V_{12}^{(1)\ast }(q+i)(\widetilde{x}_{j}-\nu _{2}) \over \omega _{k}-\varepsilon _{4}+\varepsilon _{f}-\widetilde{x}_{j}}}%
+ \\ 
+%
{\displaystyle{C_{3}V_{34}^{(2)\ast }\left[ (\widetilde{x}_{j}+\varepsilon _{2}-\varepsilon _{4}-\nu _{1})(\widetilde{x}_{j}+\varepsilon _{2}-\varepsilon _{4}+\Delta _{2}-\frac{i\Gamma _{2}}{2})-\left| V_{12}^{(1)}\right| ^{2}\right]  \over \omega _{k}-\varepsilon _{4}+\varepsilon _{f}-\widetilde{x}_{j}}}%
{\Large )}
\end{array}
\end{array}
\label{3.7}
\end{equation}

where $C_{1}(0)=C_{1},C_{3}(0)=C_{3}$ $(C_{2}(0)=C_{4}(0)=C_{f{\bf k,}%
\lambda }(0)=0)$, $\Delta _{2},$ $\Gamma _{2}$ and $\Delta _{4},$ $\Gamma
_{4}$ are the shifts and widths of levels $2$ and $4$ defined as in
preceeding section, $q$ is the Fano parameter\cite{Fano}

\begin{equation}
q=\frac{2}{\sqrt{\Gamma _{2}\Gamma _{4}}}P\sum\limits_{{\bf k},\lambda }%
\frac{\beta _{f2}({\bf k},\lambda )\beta _{f4}({\bf k},\lambda )}{\omega
_{k}+\varepsilon _{f}-\varepsilon _{2}}\approx \frac{2}{\sqrt{\Gamma
_{2}\Gamma _{4}}}P\sum\limits_{{\bf k},\lambda }\frac{\beta _{f2}({\bf k}%
,\lambda )\beta _{f4}({\bf k},\lambda )}{\omega _{k}+\varepsilon
_{f}-\varepsilon _{4}}
\end{equation}
(the levels 2 and 4 are supposed to be close having nearly equal Fano
parameters, and the $\beta $'s are taken to be real). The quantities $x_{j}$
and $\widetilde{x}_{j}$ ($j=1,2,3,4$) are the roots of the fourth order
polynomials $f(x)$ and $\widetilde{f}(x)$,respectively,

\begin{equation}
\begin{array}{c}
f(x)=\left[ (x-\nu _{1})(x+\Delta _{2}-\frac{i\Gamma _{2}}{2})-\left|
V_{12}^{(1)}\right| ^{2}\right] \times \\ 
\times \left[ (x-\varepsilon _{2}+\varepsilon _{4}-\nu _{2})(x-\varepsilon
_{2}+\varepsilon _{4}+\Delta _{4}-\frac{i\Gamma _{4}}{2})-\left|
V_{34}^{(2)}\right| ^{2}\right] - \\ 
-\frac{\Gamma _{2}\Gamma _{4}}{4}(q+i)^{2}(x-\nu _{1})(x-\varepsilon
_{2}+\varepsilon _{4}-\nu _{2}) \\ 
\widetilde{f}(x)=\left[ (x-\nu _{2})\left( x+\Delta _{4}-\frac{i\Gamma _{4}}{%
2}\right) -\left| V_{34}^{(2)}\right| ^{2}\right] \times \\ 
\times \left[ (x+\varepsilon _{2}-\varepsilon _{4}-\nu _{1})(x+\varepsilon
_{2}-\varepsilon _{4}+\Delta _{2}-\frac{i\Gamma _{2}}{2})-\left|
V_{12}^{(1)}\right| ^{2}\right] - \\ 
-\frac{\Gamma _{2}\Gamma _{4}}{4}(q+i)^{2}(x-\nu _{2})(x+\varepsilon
_{2}-\varepsilon _{4}-\nu _{1}).
\end{array}
\label{3_14}
\end{equation}

It can be shown that all the poles determined by these roots are in the
upper half-plane and satisfy the inequality

\begin{equation}
0<%
\mathop{\rm Im}%
(x_{j},\widetilde{x}_{j})<\frac{\Gamma _{2}+\Gamma _{4}}{2}.
\end{equation}

This is the main result of this section; the expression (\ref{3.7})
determines the fluorescence spectrum of the five-level atom under study.

We will analize this spectrum in two special cases where i) atom is
initially in the state 1 ($C_{1}=1,C_{3}=0$) and ii) the atom is initially
prepared in the symmetric or antisymmetric superposition state ($%
C_{1}=C_{2}=\pm 1/\sqrt{2})$.

In case where the atom is initially in the state 1, by substituting $%
C_{1}=1,C_{3}=0$ into (\ref{3.7}) we obtain that the destructive
interference may take place if ${\bf d}_{f2}$ and ${\bf d}_{f4}$ are
parallel or antiparallel, but this destructive interference is not complete
and there is no dark line in the spectrum of spontaneous transition.

Consider the case where the atom is initially in a symmetric or
antisymmetric superposition state , i.e., $C_{1}=C_{3}=\pm 1/\sqrt{2}$.
Substitution of these values into the expression (\ref{3.7}) demonstrates
that when the dipole moments ${\bf d}_{f2}$ and ${\bf d}_{f4}$ are
perpendicular to each other, the interference term disappears and the
intensity of the spontaneous emission is equal to the sum of the intensities
of the separate transitions. When the dipole moments are antiparallel, the
levels 2 and 4 are close $(\varepsilon _{2}\approx \varepsilon _{4})$, the
matrix elements $V_{12}$ and $V_{34}$ are approximately equal $%
(V_{12}\approx V_{34})$, and $d_{f2}\approx d_{f4,\text{ }}$ $\Delta
_{2}\approx \Delta _{4},$ $\Gamma _{2}\approx \Gamma _{4},$ $\nu _{1}\approx
\nu _{2}$, complete destructive interference takes place throughout the
spectrum and the intensity of spontaneous emission becomes negligible. When
the dipole moments are parallel or antiparallel, but the conditions above
are not met, constructive or destructive interference takes place at certain
frequencies depending on the sign of the interference term. Figs.3a and 4a
show this peculiarity (for the parallel case) for the values of the Fano
parameter equal to 1 and 5, respectively. For the perpendicular dipole
moments Figs. 3b and 4b demonstrate the absence the absence of the dark line
for the same values of the Fano parameter.

\section{Conclusion}

We have investigated spontaneous emission of atoms under the action of laser
radiation. We showed that the external electromagnetic field affects
essentially the spectrum of spontaneous emission by stimulating interference
between different decay channels.

In a four-level system (Fig.1a) without resonant laser radiation no
interference occurs between decay channels to levels $3$ and $4$.{\it \ }%
However{\it , }when the laser radiation is switched on and there is no
fourth level, the Autler-Townes effect is known to cause interference and
appearance of a dark line in the fluorescence spectrum \cite{Zhu} depending
on the initial conditions.

We show that if an additional level, 4, is present providing another decay
channels, this interference may partially or completely be destroyed
depending on the values of parameters. We also show that if the atom is
prepared initially in a superposition of upper states, destructive
interference occurs, under specific conditions, in both decay channels
separately, and a dark line appeares at a certain frequency.

We then show that in a five-level system where the upper states of two
two-level systems driven by two lasers can decay to the same, fifth, level,
interference between the decay channels takes place only in classical
driving fields (no interference if the fields are quantized and described in
the Fock representation) and the result of this interference depends
strongly on the mutual orientation of the dipole moment matrix elements of
the spontaneous transitions, as well as, again, on the initial conditions.

We, specifically, show that when the dipole moments of spontaneous
transitions are perpendicular to each other, the interference term
disappears and the intensity of the spontaneous emission is equal to the sum
of the intensities of the separate transitions. But when these dipole
moments are parallel or antiparallel, the destructive or constructive
interference takes place depending on the sign of the interference term. In
case where the atom is initially in the state 1, the destructive
interference is not complete and dark line is absent in the spectrum of
spontaneous emission. When the atom is initially in a superposition state, a
complete destructive interference takes place under certain conditions and
the intensity of spontaneous emission becomes very small throughout the
spectrum, while if these conditions are not met, constructive or destructive
interference takes place at certain frequencies depending on the sign of the
interference term.{\it \ }

So, the presence of additional levels enabling alternative decay channels
may modify significantly the spectrum of fluorescence from atomic media.{\it %
\ }

\section{Acknowledgement}

The authors are grateful to Prof. M.L. Ter-Mikayelyan and Dr. R.G. Unanyan
for helpful discussions$.$

The work has been performed in frames of the Project INTAS 99-00019 and the
Armenian Government grant N1350.

Fig.1. Energy-level diagrams of the four-level (a) and five-level (b) atomic
systems.

Fig. 2. The spontaneous emission spectra $S(\omega _{k})$ for $\varepsilon
_{2}-\varepsilon _{3}-\omega +\Delta _{2}=2$ $,$ $\varepsilon
_{1}-\varepsilon _{3}+\Delta _{1}=2.5,$ $\varepsilon _{2}-\varepsilon
_{4}+\Delta _{2}=3.5$, $\varepsilon _{1}-\varepsilon _{4}+\omega +\Delta
_{1}=4$, $\Gamma _{1}=\Gamma _{2}=\Gamma =1$ and (a) $\Omega =$ $0.5$, (b) $%
\Omega $ $=$ $2.5$, (c) $\Omega $ $=3.0$, (d) $\Omega $ $=$ $4.0$. All
parameters are in units of $\Gamma $.

Fig.3. The spontaneous emission spectra $S(\omega _{k})$ for the value of
Fano parameter $q$ $=1$ and following values of other parameters: $V_{12}=1$%
, $\varepsilon _{2}=\varepsilon _{4}=8$ , $\varepsilon _{f}=1$ , $\Delta
_{2}=\Delta _{4}=0$ , $\Gamma _{2}=\Gamma _{4}=1$ , $V_{34}=1.5$ . All
parameters are in units of $\Gamma $. In case (a) the transition dipole
moments are parallel, while in case (b) they are perpendicular.

Fig.4. The same as in Fig.3, but with $q$ $=5$.

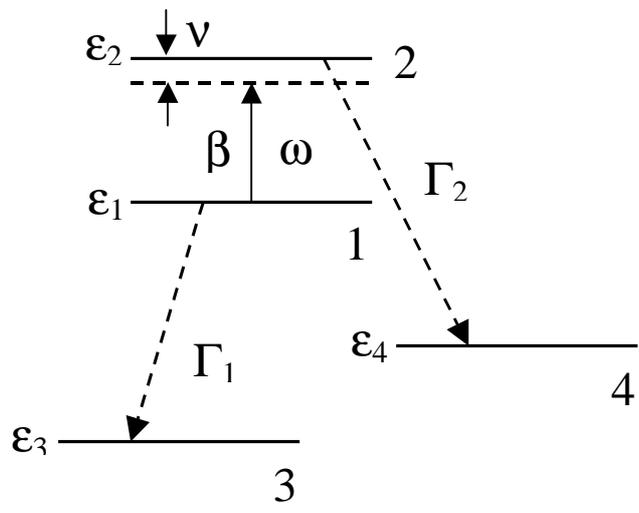
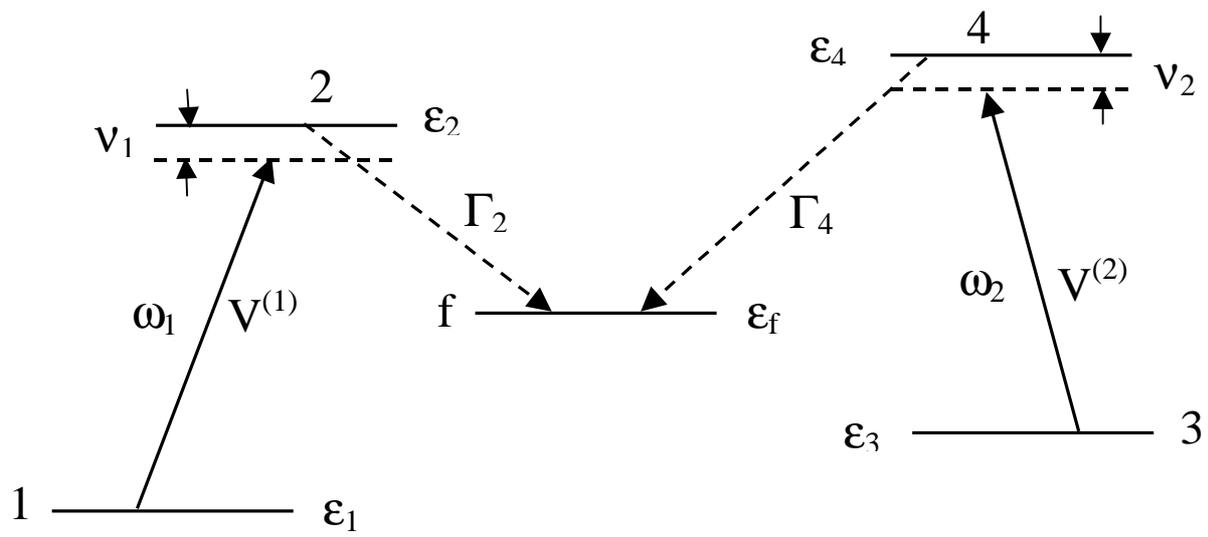
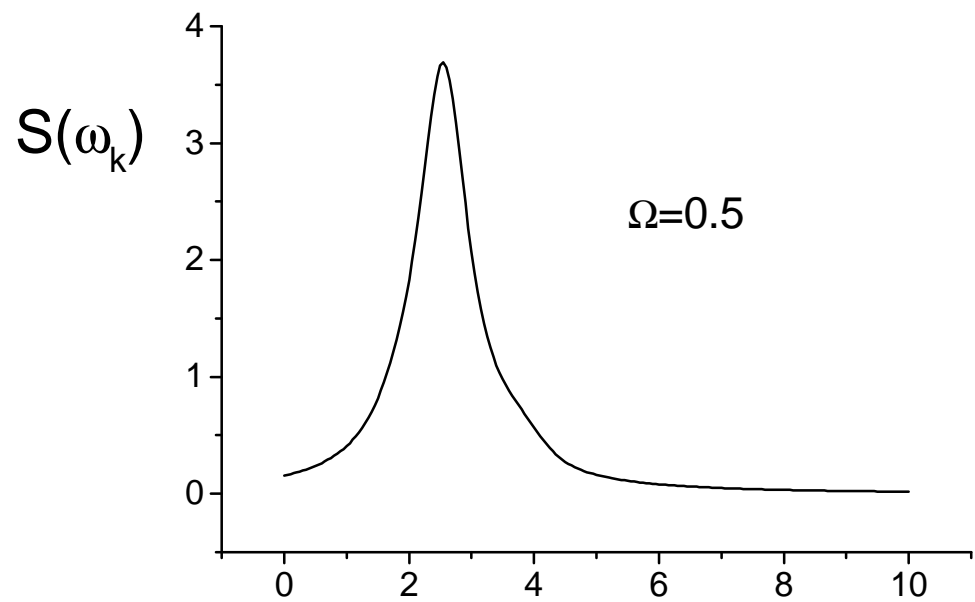
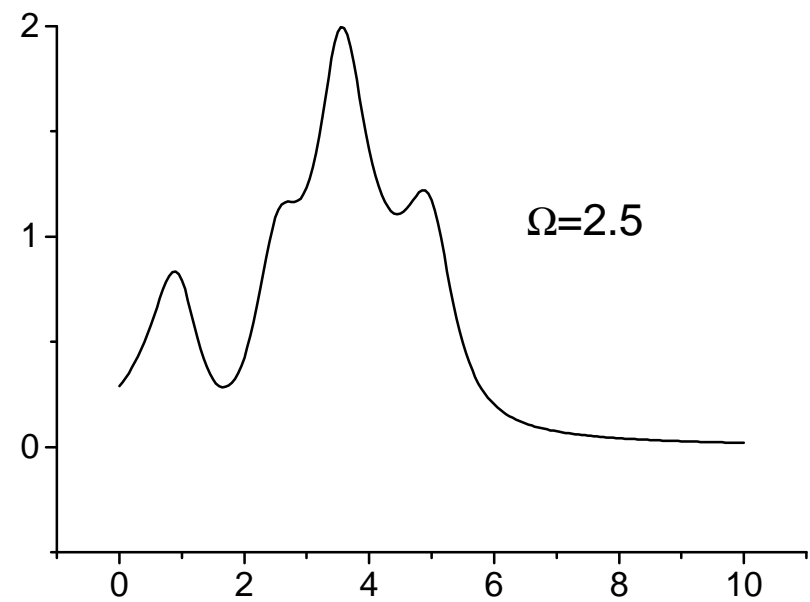
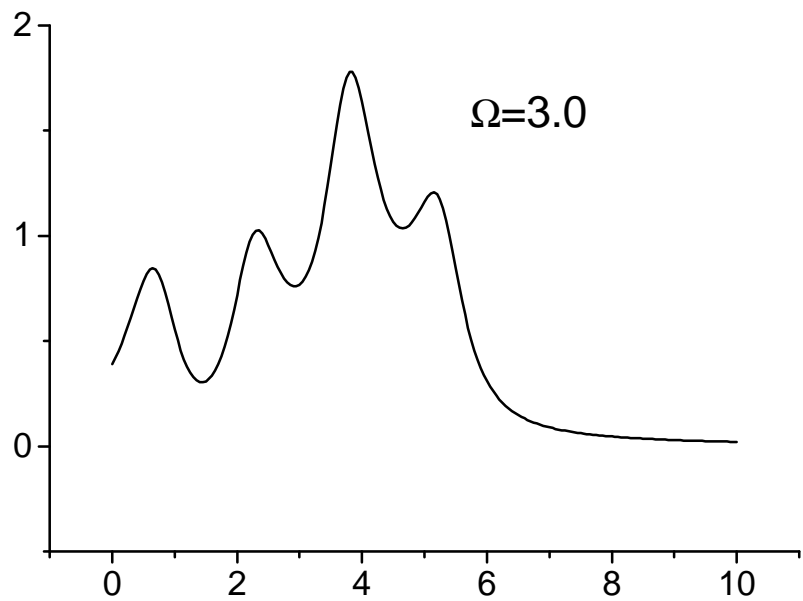
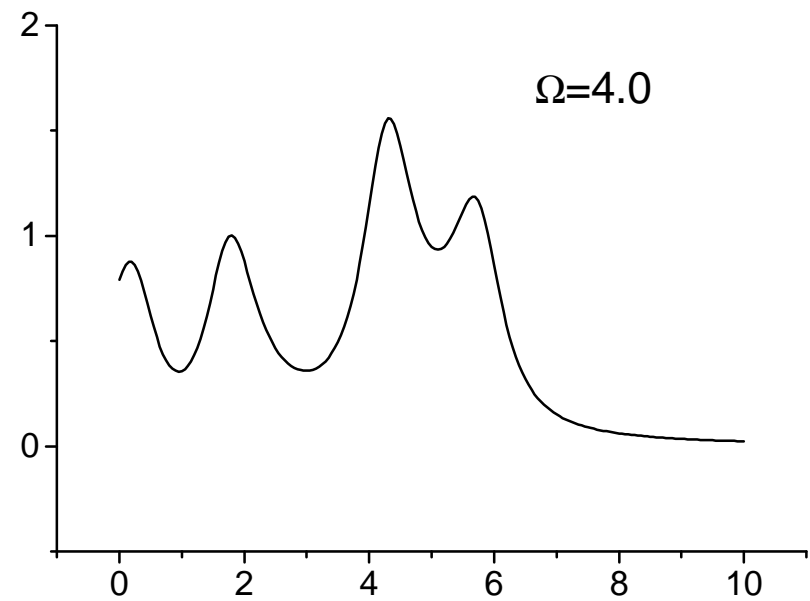
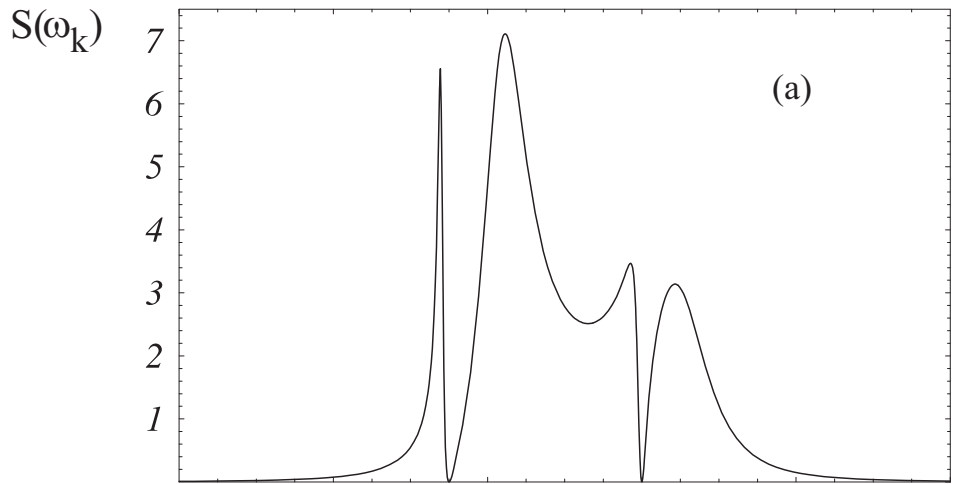
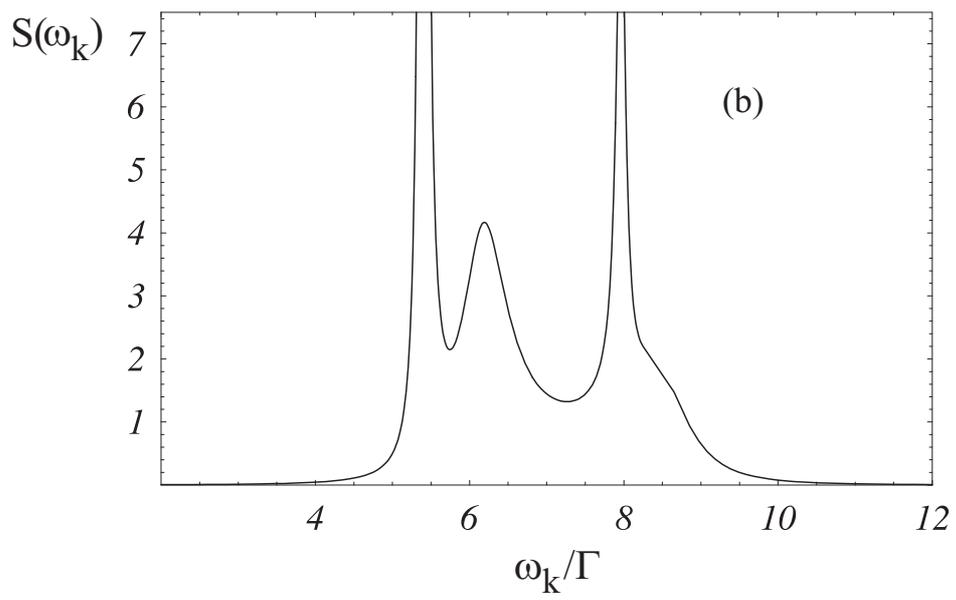
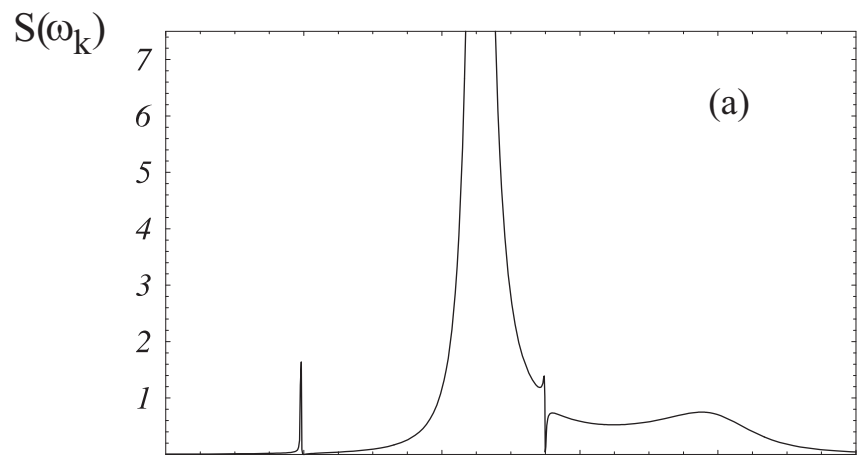
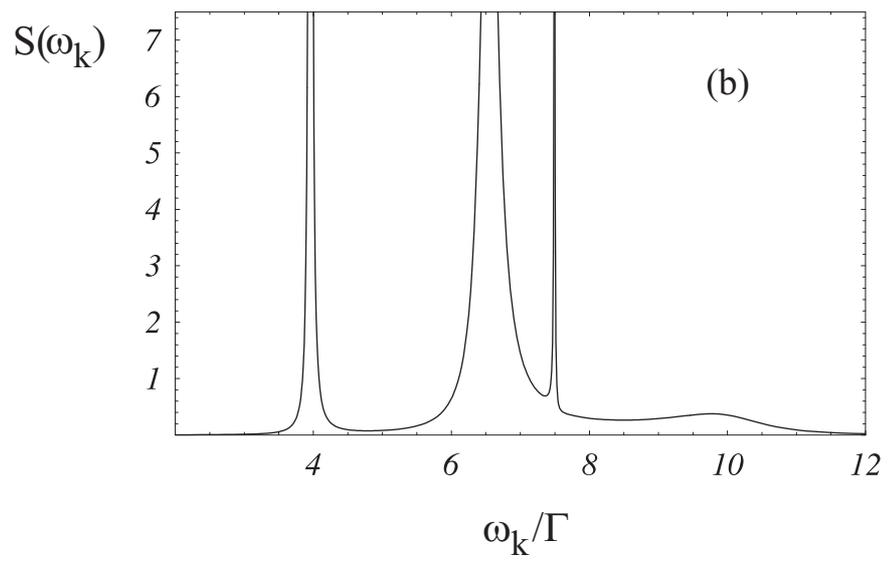
\end{document}